\documentclass[a4paper]{amsart}
\usepackage{latexsym}
\usepackage{amsmath}
\usepackage{amssymb}
\usepackage{amsfonts}

\newcounter{smallarabics}
\newenvironment{arabicenumerate}
{\begin{list}{{\normalfont\textrm{(\arabic{smallarabics})}}}
  {\usecounter{smallarabics}\setlength{\itemindent}{0cm}
   \setlength{\leftmargin}{5ex}\setlength{\labelwidth}{4ex}
   \setlength{\topsep}{0.75\parsep}\setlength{\partopsep}{0ex}
   \setlength{\itemsep}{0ex}}}
{\end{list}}

\newcounter{smallroman}

\newcommand{\ben}{\begin{arabicenumerate}}  
\newcommand{\een}{\end{arabicenumerate}}

\def\init{\setcounter{equation}{0}}

\newtheorem{theoreme}{Theorem }[section]
\newtheorem{proposition}[theoreme]{Proposition}
\newtheorem{lemma}[theoreme]{Lemma}
\newtheorem{definition}[theoreme]{Definition}
\newtheorem{corollary}[theoreme]{Corollary}
\newtheorem{remark}[theoreme]{Remark}
\newtheorem{example}[theoreme]{Example}
\newcommand{\beq}{\begin{equation}}
\newcommand{\eeq}{\end{equation}}

\newcommand{\bex}{\begin{example}}
\newcommand{\eex}{\end{example}}
\def\bel{\begin{lemma}}
\def\eel{\end{lemma}}
\def\bet{\begin{theoreme}}
\def\eet{\end{theoreme}}
\def\bed{\begin{definition}}
\def\eed{\end{definition}}
\def\ber{\begin{remark}}
\def\eer{\end{remark}}

\def\rr{{\mathbb R}}
\def\zz{{\mathbb Z}}
\def\cc{{\mathbb C}}
\def\nn{{\mathbb N}}

\def\part{{\rm par}}

\def\slim{{\rm s-}\lim}

\def\bar{\overline}

\def\cinf{C^\infty}
\def\c0inf{C_0^\infty}

\def\s{{\rm s}}

\def\proof{
\noindent{\bf Proof.}\ \ }

\def\ch{{\mathfrak h}}

\def\cZ{{\mathcal Z}}
\def\cY{{\mathcal Y}}

\def\i{{\rm i}}
\def\ad{{\rm ad}}

\def\Dom{{\rm Dom}}

\def\Ker{{\rm Ker}}

\def\qed{$\Box$\medskip}

\def \p{ \partial}

\def\12{\frac{1}{2}}
\def\14{\frac{1}{4}}

\def\ad{{\rm ad}}
\def\e{{\rm e}}

\def\d{{\rm d}}

\def\bbbone{{\mathchoice {\rm 1\mskip-4mu l} {\rm 1\mskip-4mu l}
{\rm 1\mskip-4.5mu l} {\rm 1\mskip-5mu l}}}
\def\one{\bbbone}
\def\cH{{\mathcal H}}

\def\ii{{\rm j}}

\def\Ker{{\rm Ker}}

\def\coinf{C_0^\infty}

\def \p{ \partial}

\def\12{\frac{1}{2}}

\def\ad{{\rm ad}}
\def\e{{\rm e}}

\def\d{{\rm d}}

\def\cH{{\mathcal H}}

\def\bep{\begin{proposition}}
\def\eep{\end{proposition}}

\def\b{{\rm b}}

\def\Op{{\rm Op}}
\def\s{{\rm s}}

\newcommand{\mat}[4]{\left[\begin{array}{cc}#1 &#2  \\ #3 &#4 \end{array}\right]}

\def\CARal{{\rm C\hskip 0.25 em \hbox{\raise 1.72 ex 
\hbox{$\scriptscriptstyle\rm al$}\kern -0.57 em A}R}}

\def\otimesal{\mathop{\hbox{\raise 1.5 ex
  \hbox{$\scriptscriptstyle\rm al$}
\kern -0.92 em \hbox{$\otimes$}}}}
\def\oplusal{\mathop{\hbox{\raise 1.5 ex
  \hbox{$\scriptscriptstyle\rm al$}
\kern -0.92 em \hbox{$\oplus$}}}}
\def\Gammal{\hbox{\raise 1.68 ex 
\hbox{$\scriptscriptstyle\rm al$}\kern -0.50 em $\Gamma$}}
\def\Bal{\hbox{\raise 1.68 ex 
\hbox{$\scriptscriptstyle\rm  al$}\kern -0.50 em $B$}}
\def\CARal{{\rm C\hskip 0.25 em \hbox{\raise 1.72 ex 
\hbox{$\scriptscriptstyle\rm al$}\kern -0.57 em A}R}}

\begin{document}
\title[Spectral and scattering theory of charged $P(\varphi)_2$ models]{Spectral and scattering theory \\ of space-cutoff charged $P(\varphi)_{2}$
models}
\author[C. G\'erard]{C. G\'erard \\ Laboratoire de math\'ematiques, Universit\'e de Paris XI,\\
91\,405 Orsay Cedex France}\date{July 2009}
\maketitle
\begin{abstract}
We consider in this paper space-cutoff charged $P(\varphi)_{2}$ models arising from
the quantization of the non-linear charged Klein-Gordon equation:
\[
(\p_{t}+\i V(x))^{2}\phi(t, x)+ (-\Delta_{x}+ m^{2})\phi(t,x)+
g(x)\p_{\overline{z}}P(\phi(t,x), \overline{\phi}(t,x))=0, 
\]
where  $V(x)$ is an electrostatic potential, $g(x)\geq 0$  a  space-cutoff and $P(\lambda,
\overline{\lambda})$  a real bounded below polynomial.
We discuss various ways to quantize  this equation, starting from
different CCR representations. After describing the construction of
the interacting Hamiltonian $H$ we study its spectral and scattering
theory. We describe the essential spectrum of $H$, prove the existence
of asymptotic fields and of wave operators, and finally prove the {\em asymptotic
completeness} of wave operators. These results are similar to the case
when $V=0$.
\end{abstract}
\section{Introduction}\label{sec0}\init
\subsection{Charged Klein-Gordon equations}\label{sec0.1}
Let us consider the charged Klein-Gordon equation:
\beq
\label{e0.1}
(\p_{t}+\i V(x))^{2}\phi(t, x)+ (-\Delta_{x}+ m^{2})\phi(t,x)
=0, 
\eeq
where $\phi: \rr_{t}\to L^{2}(\rr^{d}; \cc)$, $m>0$ is the mass. The
equation (\ref{e0.1})
describes a charged  field minimally coupled to a  external
electrostatic field given by the potential $V$. As is well known, after
introducing the $\varphi$ and $\pi$ fields by
\[
\varphi(t)= \phi(t), \ \pi(t)= \p_{t}\phi(t)+\i V\phi(t),
\]
one can interpret (\ref{e0.1}) as a  Hamiltonian system on the
symplectic space
\[
\cY=\{y=(\pi, \varphi)\ : \ \pi, \varphi\in L^{2}(\rr^{d}, \cc)\}
\]
equipped with the (complex) symplectic form
\[
(\pi, \varphi)\omega (\pi', \varphi')= \int_{\rr^{d}} \overline{\pi}(x)\varphi'(x)-\overline{\varphi}(x)\pi'(x)\d x.
\]
for
the classical Hamiltonian
\[
\begin{array}{rl}
&h_{V}(\pi, \varphi)\\[2mm]
=&\int_{\rr^{d}}\overline{\pi}(x)\pi(x)\d x+ \int_{\rr^{d}}
\overline{\nabla_{x}\varphi}(x)\cdot\nabla_{x}\varphi(x)+
m^{2}\overline{\varphi}(x)\varphi(x)\d x\\[2mm]
&+\i\int_{\rr^{d}}
\overline{\varphi}(x)V(x)\pi(x)-\overline{\pi}(x)V(x)\varphi(x)\d x
\end{array}
\]
In order to obtain a {\em stable quantization} of (\ref{e0.1}), ie a
CCR representation of $(\cY, \omega)$ in a Hilbert space $\cH$ with
the property that  the
time evolution is implemented by a positive Hamiltonian, it is
necessary that the classical Hamiltonian $h_{V}(\pi, \varphi)$ is {\em
positive}. If this is the case,  one can equip $\cY$ with a {\em K\"{a}hler
structure}, ie a complex structure $\ii$ such that
\[
(y| y')_{\rm dyn}:= y\omega\ii y'+ \i y\omega y'
\]
is a scalar product on $\cY$. The completion  of the
pre-Hilbert space $(\cY, \ii, (\cdot|\cdot)_{\rm dyn})$, denoted by
$\cZ$ is called the {\em one-particle space}. The stable
quantization is then the Fock representation on the bosonic Fock space
$\Gamma_{\s}(\cZ)$, and the time evolution is unitarily implemented by
the group $\e^{\i t H_{V}}$, where $H_{V}= \d\Gamma(h_{V})$ is a
second quantized Hamiltonian.

An alternative quantization is obtained by considering first the
Klein-Gordon equation (\ref{e0.1}) for $V(x)\equiv 0$. 
Let us  denote by
$\ii_{0}$ (resp. $\cZ_{0}$) the associated complex structure  (resp.
one-particle space). As is well known, $\cZ_{0}$ can be unitarily
identified with $L^{2}(\rr^{d})\oplus L^{2}(\rr^{d})$.  

The dynamics for $V=0$ is unitarily implemented
by $\e^{\i t H_{0}}$ on the Fock space $\Gamma_{\b}(\cZ_{0})$, for
$H_{0}= \d\Gamma(\omega)$,
where  $\omega= \epsilon\oplus \epsilon$ is the one-particle energy
and $\epsilon=(-\Delta_{x}+ m^{2})^{\12}$.

One can then try to implement the dynamics for $V\neq 0$ by
considering the Fock representation on $\Gamma_{\s}(\cZ_{0})$ and by
giving a meaning to the formal expression:
\[
H= \d\Gamma(\omega)+ \i\int_{\rr^{d}}
\overline{\varphi}(x)V(x)\pi(x)-\overline{\pi}(x)V(x)\varphi(x)\d x,
\]
where $\varphi(x)$, $\pi(x)$ are the {\em quantized} $\varphi$ and $\pi$
fields.
Note that the two CCR representations above are in general not unitarily
equivalent.

It turns out that it is possible to give a meaning to $H$,in {\em one space dimension} ($d=1$),
provided the potential $V$ is small enough as we will see in Sect.
\ref{sec3}.

\subsection{Non-linear perturbations}\label{sec0.2}

We assume now that $d=1$.  Let us  fix a positive space cutoff
function $g:\rr\to
\rr^{+}$, decreasing fast enough at infinity and a bounded below real
potential $P(\lambda, \overline{\lambda})$. We consider now the
{\em non-linear} charged Klein-Gordon equation:
\beq
(\p_{t}+\i V(x))^{2}\phi(t, x)+ (-\Delta_{x}+ m^{2})\phi(t,x)+
g(x)\p_{\overline{z}}P(\phi(t,x), \overline{\phi}(t,x))=0. 
\label{e0.3}
\eeq
The usual procedure to quantize (\ref{e0.3}) is to start from a
quantization of (\ref{e0.1}) (ie (\ref{e0.3}) for $g(x)\equiv 0$), leading to the
Hamiltonians $H_{V}$ or $H_{0}$ (depending on the choice of the CCR
representation), and to implement the interacting dynamics by giving a
meaning to  either:
\beq\label{e0.4}
H_{V}+ \int_{\rr}g(x)P(\varphi(x), \overline{\varphi}(x))\d x,
\eeq
or:
\beq\label{e0.5}
H_{0}+ \i\int_{\rr}
\overline{\varphi}(x)V(x)\pi(x)-\overline{\pi}(x)V(x)\varphi(x)\d x+\int_{\rr}g(x)P(\varphi(x), \overline{\varphi}(x))\d x.
\eeq
The choice (\ref{e0.4}) seems difficult, because both the one-particle
energy $h_{V}$ and the $\varphi$, $\pi$ fields are not very explicit
in the Fock representation for the complex structure $\ii$.

In this paper we will adopt the choice (\ref{e0.5}). 

The associated
Hamiltonian will be constructed in Sect. \ref{sec3}. We will show that
if $|\lambda|<\lambda_{\rm quant}$, where the constant $\lambda_{\rm
quant}$ is defined in (\ref{def-de-lambda}), the formal expression
\[
H:=H_{0}+ \i\lambda\int_{\rr}
\overline{\varphi}(x)V(x)\pi(x)-\overline{\pi}(x)V(x)\varphi(x)\d x+
\int_{\rr}g(x):\!P(\varphi(x), \overline{\varphi}(x))\!:\d x,
\]
is well defined as a bounded below selfadjoint operator.

The rest of the paper is devoted to the spectral and scattering theory
of $H$, which is studied in Sect. \ref{sec4}. We will use the results of \cite{GP}, where an abstract class
of QFT Hamiltonians are considered, so most of our task is to prove that
our Hamiltonian $H$ satisfies the abstract hypotheses of \cite{GP}.
This will be done in Subsect. \ref{sec4.1}.

The first result is the {\em HVZ theorem}, describing the essential
spectrum of $H$. We obtain that
\[
\sigma_{\rm ess}(H)= [{\rm inf}\sigma(H)+ m, +\infty[,
\]
which implies that $H$ has a ground state.

The second results deal with the {\em scattering theory} of $H$, which
is formulated in terms of {\em asymptotic fields}. These are
(formally) defined as the limits:
\[
\lim_{t\to \pm\infty}\e^{\i tH}\phi(\e^{\i t\omega}F)\e^{-\i tH}=:
\phi^{\pm}(F), \ F\in L^{2}(\rr)\oplus L^{2}(\rr).
\]
It follows then from the stability condition $|\lambda|<\lambda_{\rm
quant}$ and abstract arguments that the two asymptotic CCR
representations
\[
F\mapsto \phi^{\pm}(F)
\]
are of {\em Fock type}, ie unitarily equivalent to a sum of Fock
representations. 

The main problem of scattering theory is now to
identify the spaces of {\em asymptotic vacua}, ie  the spaces of
vectors annihilated by all {\em asymptotic annihilation operators}
$a^{\pm}(F)$. Applying the abstract results of \cite{GP}, we show that
the asymptotic vacua coincide with the {\em bound states} of $H$. This
result, called the {\em asymptotic completeness} of wave operators, is
the main result of this paper.

\subsection{Notation}\label{sec0.3}
In this subsection we collect some useful notation and results.

\medskip

{\it Scales of Hilbert spaces}

\medskip

If  $\ch$ is  a Hilbert space  and $\epsilon$ a linear
operator on $\ch$, its domain will be denoted by $\Dom \epsilon$.  The
closure of a closeable operator $a$  will be denoted by $a^{\rm cl}$.

If $\epsilon$ is selfadjoint,  we write $\epsilon>0$ if $\epsilon\geq 0$ and
$\Ker \epsilon=\{0\}$. If $\epsilon>0$ and $s\in \rr,$
$\epsilon^{s}$ is well defined as a selfadjoint operator and we denote
by $\epsilon^{s}\ch$ the completion of $\Dom \epsilon^{-s}$ for the
norm $\|\epsilon^{-s}h\|$. Clearly $\epsilon^{s}\ch$ are Hilbert
spaces and $\epsilon^{t}$ is isometric from $\epsilon^{s}\ch$ to
$\epsilon^{s+t}\ch$.

\medskip

{\it Fourier transform}

\medskip

Let   $\ch= L^{2}(\rr)$. We denote by ${\mathcal F}$
the unitary Fourier transform:
\[
{\mathcal F}u(k)= (2\pi)^{-1/2}\int_{\rr}\e^{-\i k\cdot x}u(x)\d x.
\] 
We denote also by $\widehat{f}$ the usual Fourier transform of $f$:
\[
\widehat{f}(k)=\int_{\rr}\e^{-\i k\cdot x}u(x)\d x,
\]
so that if $V$ is the  operator of multiplication by the function $V$ one
has
\beq\label{e1.00}
{\mathcal F}V{\mathcal F}^{-1}u(k_{1})=
(2\pi)^{-1}\int_{\rr}\widehat{V}(k_{1}- k_{2})u(k_{2})\d k_{2}.
\eeq
If $\epsilon= (-\Delta_{x}+ m^{2})^{\12}$
for $m>0$ then $\epsilon^{s}L^{2}(\rr)$ is equal to the Sobolev
space $H^{-s}(\rr)$ with the norm
\[
\|f\|^{2}_{H^{-s}(\rr)}= \int_{\rr}(k^{2}+ m^{2})^{-s}|{\mathcal
F}u(k)|^{2}\d k.
\]

\medskip

{\it Pseudodifferential calculus}

\medskip

Set $\langle x\rangle=(1+ x^{2})^{\12}$. For $m\in \rr$ we will denote by $S^{m}(\rr)$ the space
\[
S^{m}(\rr)= \{f\in \cinf(\rr)\ : \ |f^{(\alpha)}(x)|\leq C_{n}\langle
x\rangle^{m-
\alpha}, \ \alpha\in \nn\}.
\]
For $m, p\in \rr$ we denote by $S^{m, p}(\rr^{2})$ the space
\[
S^{m, p}(\rr^{2})= \{f\in \cinf(\rr^{2})\ : \
|\p_{x}^{\alpha}\p_{k}^{\beta}f(x, k)|\leq C_{\alpha, \beta}\langle
x\rangle^{m-\alpha}\langle k\rangle^{p-\beta}, \ \alpha, \beta\in \nn\}.
\]
If $a\in S^{m, p}(\rr^{2})$, we denote by $\Op^{\rm w}(a)= a^{\rm
w}(x, D_{x})$ the {\em Weyl quantization} of $a$, defined as:
\[
\Op^{\rm w}(a)u(x)= (2\pi)^{-1}\int\e^{\i (x-y)\cdot k}a(\frac{x+
y}{2}, k)u(y)\d y \d k,
\]
as an operator on ${\mathcal S}(\rr)$, where ${\mathcal S}(\rr)= \bigcap_{m\in
\rr}S^{m}(\rr)$ is  the Schwartz class. 

The operator $\Op^{\rm w}(a)$ is bounded on $L^{2}(\rr)$ if $a\in
S^{m, p}(\rr^{2})$ for $m, p\leq
0$, and belongs to the Hilbert-Schmidt class iff $a\in
L^{2}(\rr^{2})$. One has then
\beq\label{e1.0a}
\|\Op^{\rm w}a\|^{2}_{\rm HS}= \frac{1}{2\pi}\int_{\rr^{2}}|a(x,
k)|^{2}\d x
\d k.
\eeq
\section{Charged Klein-Gordon equation}
\label{sec1}\init
In this section we detail the arguments given in Subsect.
\ref{sec0.1}.
The results of this section are standard,  they can be found for
example in  Palmer \cite{Pa}. For simplicity we consider the one
dimensional case, although  the results of Subsect. \ref{sec1.2} hold
in any space dimension. 
\subsection{Charged Klein-Gordon equation as a Hamilton
equation}\label{sec1.2}
Let $m>0$ and $V: \rr\to \rr$ a real   measurable potential such that 
\beq\label{e1.0}
V, \ \nabla_{x}V\in L^{\infty}(\rr).
\eeq
We consider the Cauchy problem for the 
 charged  Klein-Gordon equation:
\beq
\left\{\begin{array}{l}
(\p_{t}+\i V(x))^{2}\phi(t, x)+ (-\Delta_{x}+ m^{2})\phi(t,x)=0,\\[2mm]
\phi(0,x)= \varphi(x), \\[2mm]
\ \p_{t}\phi(0, x)+ \i V(x) \phi(0,x)= \pi(x), 
\end{array}\right.
\label{e1.1}
\eeq
where $\phi: \rr\to L^{2}(\rr; \cc)$, describing a charged scalar field
of mass $m$ minimally coupled to the electrostatic potential $V$.

Note that (\ref{e1.1}) is invariant under {\em time-reversal}, ie if
$\phi(t,x)$ is a solution, so is $\overline{\phi}(-t, x)$. In terms of
Cauchy data, time-reversal becomes the involution:
\beq\label{invol}
\kappa: (\pi, \varphi)\mapsto (-\overline{\pi}, \overline{\varphi}).
\eeq
Let us set 
\[
\varphi(t):= \phi(t), \ \pi(t)= \p_{t}\phi(t)+ \i V \phi(t),
\]
and
\[
\cY=\{y=(\pi, \varphi) \ : \ \pi, \varphi\in L^{2}(\rr)\}.
\]
We transform (\ref{e1.1}) into the first order evolution equation on
$\cY:=L^{2}(\rr)\oplus L^{2}(\rr)$:
\[
\left[\begin{array}{c}
\pi(t)\\\varphi(t)
\end{array}\right]= r_{t}\left[\begin{array}{c}
\pi(0)\\\varphi(0)
\end{array}\right]
\]
Formally we have,
\beq\label{e1.2}
\p_{t}\left[\begin{array}{c}
\pi(t)\\\varphi(t)
\end{array}\right]= \left[\begin{array}{cc}
-\i V&-\epsilon^{2}\\\one &-\i V
\end{array}\right]\left[\begin{array}{c}
\pi(t)\\\varphi(t)
\end{array}\right],
\eeq
for $\epsilon= := (-\Delta_{x}+ m^{2})^{\12}$.

If we equip $\cY$ with the (sesquilinear) anti-symmetric form:
\[
(\pi, \varphi)\omega(\pi',\varphi')= (\pi|\varphi')-(\varphi|\pi'),
\]
and the Hamiltonian:
\beq\label{e1.3}
h_{V}(\pi, \varphi)= \|\pi\|^{2}+ \|\epsilon \varphi\|^{2}+ \i(\varphi|
V\pi)-\i(\pi|V\varphi),
\eeq
we see that  (\ref{e1.2}) are the associated Hamilton equations. If we prefer
to forget the complex structure of $\cY$, we write
\beq\label{neutra}
\varphi=: \varphi_{1}+\i \varphi_{2}, \ \pi=:\pi_{1}+\i \pi_{2},
\eeq
and equip $\cY$ (as a real vector space) with the real symplectic form
${\rm Re}\omega$ and the Hamiltonian
\beq
\begin{array}{rl}
h_{V, \rr}(\pi, \varphi)=& \12 h(\pi, \varphi)\\[2mm]
=& \12\|\pi_{1}\|^{2}+ \12\|\pi_{2}\|^{2}+ \12 \|\epsilon
\varphi_{1}\|^{2}+ \12\|\epsilon\varphi_{2}\|^{2}\\[2mm]
&+(\pi_{1}| V \varphi_{2})-(\pi_{2}| V\varphi_{1}).
\end{array}
\label{e1.2bis}
\eeq
\subsection{Stable quantization}\label{sec1.3}
 A {\em stable } quantization of the 
symplectic dynamics $r_{t}$ is a CCR representation of the
symplectic space $(\cY,  \omega)$:
\[
\cY\ni y\mapsto W(y)\in U(\cH)
\]
in some Hilbert space $\cH$ such that there exists a {\em positive}
selfadjoint operator $H$ on $\cH$ implementing $r_{t}$, ie:
\[
\e^{\i tH }W(y)\e^{-\i tH}= W(r_{t}y), \ y\in \cY, \ t\in \rr.
\]
As is well known (see eg \cite{BSZ}), in order for a stable quantization to exist, it is
necessary that the classical Hamiltonian $h_{V}(\pi, \varphi)$ is
positive. The violation of the positivity of $h_{V}(\pi, \varphi)$ is connected with the so called {\em
Klein paradox}.

Let us  assume the following stronger positivity:
\beq\label{e1.5}
\pm \i\left((\varphi|V\pi)-(\pi|V\varphi)\right)\leq
\delta\left(\|\pi\|^{2}+ \|\epsilon \varphi\|^{2}\right) \ \pi\in
L^{2}(\rr), \ \varphi\in \Dom \epsilon,\hbox{ for  }0\leq
\delta<1.
\eeq
Note that 
(\ref{e1.5}) implies that  the energy norms $h_{0}(\cdot)^{\12}$ and
$h_{V}(\cdot)^{\12}$ are equivalent.

The construction of the stable quantization is then as follows:
\begin{enumerate}

\item one considers the  {\em energy space} $\cY_{\rm en}$ which is the
completion of $L^{2}(\rr)\oplus H^{1}(\rr)$ for the norm
$h_{V}(\pi, \varphi)^{\12}$;  

\item clearly $t\to r_{t}$ is 
a strongly continuous group of isometries of $\cY_{\rm en}$,
and we denote by $a$ its generator ie $r_{t}=: \e^{ta}$.   From
(\ref{e1.2}) we see that
\[
a= \left[\begin{array}{cc}
-\i V&\Delta_{x} -m^{2}\\\one &-\i V
\end{array}\right],
\]
is anti-selfadjoint on $\Dom a=H^{1}(\rr)\oplus
H^{2}(\rr)$.
Moreover from (\ref{e1.5}) we see that $\Ker a=\{0\}$.

\item we consider now the polar decomposition of $a$: 
\[
h_{V}:= (-a^{2})^{\12}, \ a=: \ii h_{V}= \ii h_{V},
\]
and we see that $\ii$ is an anti-involution (a complex structure) on
$\cY_{\rm en}$, such that $\omega\ii$ is a symmetric positive definite
form.

\item  we equip $\cY_{\rm en}$ with the complex structure $\ii$ and the
 scalar product
\[
(y_{1}| y_{2})_{\rm dyn}:= y_{1}\omega\ii y_{2}+\i y_{1}\omega y_{2}.
\]

\item denoting by $\cZ$ the completion of $\cY_{\rm en}$ for $(\cdot |\cdot)$,
we obtain a complex Hilbert space, such that $h_{V}$ extends to $\cZ$ as a
positive selfadjoint operator. The stable quantization of the charged
Klein-Gordon equation is obtained by taking the Hilbert space:
\[
\cH= \Gamma_{\s}(\cZ),
\]
where $\Gamma_{\s}(\cZ)$ is the bosonic Fock space over $\cZ$, the CCR
representation
\[
\cZ\supset\cY_{\rm en}\ni y\mapsto W(y) \in U(\cH)
\]
where $W(y)$  are the Fock Weyl operators, and the physical
Hamiltonian
\[
H= \d\Gamma(h_{V}),
\]
where $\d\Gamma(h_{V})$ is the second quantization of $h_{V}$.
\end{enumerate}
\subsection{Alternative choice of the complex structure}\label{sec1.3b}
Let us  consider the charged Klein-Gordon
equation (\ref{e1.1}) for $V=0$ and denote with the subscript $0$ the
associated objects. 

By the same procedure as above we  can equip $\cY$ with  the {\em
free}
complex structure $\ii_{0}$.  A very convenient feature of $\ii_{0}$
is that if $\cZ_{0}$ is the associated Hilbert space, then the map:
\[
U: \cZ_{0}\ni (\pi, \varphi)\mapsto (\epsilon^{-\12}\pi+\i
\epsilon^{\12} \varphi,
\epsilon^{-\12}\overline{\pi}+\i \epsilon^{\12}\overline{\varphi})\in
L^{2}(\rr^{d})\oplus L^{2}(\rr^{d})
\]
is unitary. This allows to identify $\cZ_{0}$ with an explicit Hilbert
space.
In terms of neutral  fields $\pi_{i}$, $\varphi_{i}$  the map $W$
becomes:
\beq
(\pi, \varphi)\mapsto (\epsilon^{-\12}\pi_{1}+ \i \epsilon^{\12} \varphi_{1},
\epsilon^{-\12}\pi_{2}+ \i \epsilon^{\12} \varphi_{2})\in
L^{2}(\rr^{d})\oplus L^{2}(\rr^{d}). 
\label{e1.002}
\eeq
As is well known (see eg \cite{Pa}), there
exists an invertible symplectic transformation $u$ on $\cY$ such that 
\[
\ii_{0}= u^{-1}\ii u.
\]
(This actually holds for any pair of K\"{a}hler complex structures on a
symplectic space).

Therefore  $u:\cZ\to \cZ_{0}$ and its second quantization $\Gamma(u): \Gamma_{\s}(\cZ)\to
\Gamma_{\s}(\cZ_{0})$  are unitary. The Fock representation of CCR on
$\Gamma_{\s}(\cZ)$ is unitarily equivalent to the following Bogoliubov
representation on $\Gamma_{\s}(\cZ_{0})$:
\beq\label{e1.002b}
W_{V}(f):= W_{0}(uf), \ f\in \cY,
\eeq
where $W_{0}(\cdot)$ is the Fock representation on
$\Gamma_{\s}(\cZ_{0})$. This allows to work on the more convenient
Fock space $\Gamma_{\s}(\cZ_{0})$. The positive Hamiltonian on
$\Gamma_{\s}(\cZ_{0})$ implementing the dynamics $\e^{t a}$ in the
Bogoliubov representation  $W_{V}(\cdot)$ is then
\[
\d\Gamma(h_{V}),
\]
where we still denote by $h_{V}$ acting on $\cZ_{0}$ the operator
$uh_{V}u^{-1}$.
\subsection{Quantization of the non-linear charged Klein-Gordon equation}\label{sec.1.4}

Let $P(z_{1}, z_{2})$ be a polynomial on $\cc^{2}$ such that $\cc\ni
z\mapsto P(z,
\bar{z})$ is real and bounded below. Let also $g$ a positive
function (typically $g\in \coinf(\rr)$). We consider now the non-linear
Klein-Gordon equation:
\beq
(\p_{t}+\i V(x))^{2}\phi(t, x)+ (-\Delta_{x}+ m^{2})\phi(t,x)+
g(x)\p_{\overline{z}}P(\phi(t,x), \overline{\phi}(t,x))=0. 
\label{e1.5b}
\eeq

The quantization of (\ref{e1.5b}) for $g(x)\equiv 0$, outlined in
Subsect. \ref{sec1.3b} leads to the free Hamiltonian
\[
\d\Gamma( h_{V}), \hbox{ acting on }\Gamma_{\s}(\cZ_{0}),
\]
and to the Bogoliubov representation of CCR $W_{V}(\cdot)$  defined in
(\ref{e1.002b}).

Denoting by $\phi_{V}(f)$ for $f\in \cY$ the Segal field operators
associated to the CCR representation (\ref{e1.002b}) , one sets:
\[
\varphi_{V}(x)= \phi_{V}(\delta_{x}, 0),\ x\in \rr
\]
which are the corresponding $\varphi$ fields. The natural way to
quantize (\ref{e1.5b}) is now to try to make sense of the Hamiltonian
\beq\label{hamil}
H_{V}=\d\Gamma( h_{V})+\int_{\rr}g(x)P(\varphi_{V}(x), \overline{\varphi}_{V}(x))\d
x.
\eeq
If (possibly after some Wick ordering of the interaction term), the above
Hamiltonian is well defined, one can set
\[
\phi_{V}(t, f)=\e^{\i tH_{V}}\phi_{V}(f)\e^{-\i tH},
\] 
which leads to the quantization of (\ref{e1.5b}) in the Bogoliubov
representation (\ref{e1.002b}).

The difficulty with this method is of course to make sense of $H_{V}$,
since neither the one-particle Hamiltonian $h_{V}$ nor the $\varphi$
fields $\varphi_{V}(x)$ are explicitely known. 

Actually if $V$ decays fast enough at infinity, it is possible to find
a symplectic transformation $u$ such that $uh_{V}u^{-1}$ equals the
{\em free} one-particle energy and additionally $u$ is {\em real}, ie
commutes with the time-reversal operator $\kappa$ in (\ref{invol}).
This opens the possibility to rigorously construct  the Hamiltonian
(\ref{hamil}). We plan to come back to this problem in a subsequent
paper.

An alternative way, which we will follow in this paper, is as follows:

\begin{enumerate}
\item one considers the stable quantization of (\ref{e1.1}) {\em for}
$V=0$, leading to the usual complex structure $\ii_{0}$. It is
convenient to use the  neutral fields $\pi_{i},\varphi_{i}$ $i=1, 2$
as in (\ref{neutra}), and to identify the one-particle space $\cZ_{0}$ with
$L^{2}(\rr)\oplus L^{2}(\rr)$ as in (\ref{e1.002}). 

\item the free Hamiltonian is now
\[
H_{0}=\d\Gamma( \epsilon\oplus \epsilon), \hbox{ acting on }\cH= \Gamma_{\s}(L^{2}(\rr)\oplus L^{2}(\rr)),
\]
which implements the dynamics $\e^{ta_{0}}$ in the Fock
representation for the complex structure $\ii_{0}$.

\item one sets for $x\in \rr$:
\beq\label{def-de-phi}
\begin{array}{l}
\varphi_{1}(x):=\phi\left(\epsilon^{-\12}\delta_{x}\oplus 0\right), \
\varphi_{2}(x):=\phi\left(0\oplus \epsilon^{-\12}\delta_{x}\right)\\ 
\pi_{1}(x):= \phi\left(\i
\epsilon^{\12}\delta_{x}\oplus 0\right), \ \pi_{2}(x):= \phi\left(0\oplus \i
\epsilon^{\12}\delta_{x}\right),
\end{array}
\eeq
where $\phi(f)$ are the Segal field operators. These operators are well defined  as selfadjoint operators after integration
against test functions.

\item setting with a slight abuse of notation
\[
P(\lambda_{1}, \lambda_{2}):= P(\lambda_{1}+ \i \lambda_{2},
\lambda_{1}- \i \lambda_{2}),
\]
\end{enumerate}
one tries to rigorously define as a selfadjoint operator the formal
expression:
\beq\label{e1.003}
H=\d\Gamma(\epsilon\oplus \epsilon)+ \int_{\rr}g(x) P(\varphi_{1}(x),
\varphi_{2}(x))\d x + \int_{\rr}V(x)\left(\pi_{1}(x)\varphi_{2}(x)-
\pi_{2}(x)\varphi_{1}(x)\right)\d x,
\eeq
corresponding to the hamiltonian $h_{\rr}(\pi, \varphi)$ defined in
(\ref{e1.2bis}). This will be done in Sect. \ref{sec3}.

\section{Local charge operator}\init\label{sec-charge}
In the rest of the paper we set
\[
\ch= L^{2}(\rr)\oplus L^{2}(\rr), \ \cH= \Gamma_{\s}(\ch).
\]
The elements of $\ch$ will be denoted by $F=(f_{1},f_{2})$. The
one-particle energy is
\[
\omega:=\epsilon\oplus \epsilon\hbox{ acting on }\ch,
\]
and 
\[
H_{0}:=\d\Gamma(\omega)
\]
 The (total) {\em number operator} $N$ is
\[
N:= \d\Gamma( \one\oplus \one),
\]
equal to $N_{1}+ N_{2}$, where
\[
N_{1}:= \d\Gamma(\one\oplus 0), \ N_{2}:= \d\Gamma(0\oplus \one).
\]
We will also use  the partial creation/annihilation operators
\[
a_{1}^{\sharp}(f)= a^{\sharp}(f\oplus 0) \ a_{2}^{\sharp}(f)=
a^{\sharp}(0\oplus f), \ f\in L^{2}(\rr).
\]
\subsection{Local charge operator}\label{sec2.1}
 Set
\[Q(V):=
\int_{\rr}V(x)\left(
\pi_{1}(x)\varphi_{2}(x)- \pi_{2}(x)\varphi_{1}(x)\right)\d x,
\]
where $\varphi_{i}(x)$, $\pi_{i}(x)$ are defined in
(\ref{def-de-phi}). For the moment it is only a formal expression. 

We will call $Q(V)$ a  {\em local charge operator}. 

To work with
well defined objects, we introduce the UV cutoff fields,
$\varphi_{i}^{\kappa}(x)$, $\pi^{\kappa}_{i}(x)$,  for $\kappa\gg 1$,
obtained by replacing
 $\delta_{x}$ by $F(\kappa^{-1}D_{x})\delta_{x}$ where $F\in
\coinf(\rr)$ is a cutoff function with $F(0)=1$. We denote by
$Q^{\kappa}(V)$ the cutoff charge operator, wich is for example well
defined on $\Dom N$.

\begin{lemma}\label{2.1}
Assume that
$V\in L^{\infty}(\rr)$ and $V'\in L^{\infty}(\rr)\cap
L^{2}(\rr)$. Then  there exists a constant $C$ such that:
\beq\label{e2.0c}
\|\ Q(V)(N+1)^{-1}\|\leq
C(\|V\|_{\infty}+ \|V\|_{2}+ \|V'\|_{\infty}).
\eeq
\end{lemma}
\proof
For ease of notation we will remove the UV cutoff.  To get a rigorous
proof, it suffices to put back the UV cutoff, letting $\kappa\to
+\infty$ in the various estimates.

Introducing the creation/annihilation operators $a_{i}^{\sharp}(f)$
for $i=1,2$ we have:
\[
\begin{array}{rl}
& \pi_{1}(x)\varphi_{2}(x)- \pi_{2}(x)\varphi_{1}(x)\\[2mm]
=&\frac{\i}{2}\left(a_{1}^{*}(\epsilon^{\12}\delta_{x})
-a_{1}(\epsilon^{\12}\delta_{x})\right)\left(a_{2}^{*}(\epsilon^{-\12}\delta_{x})
+a_{2}(\epsilon^{-\12}\delta_{x})\right)\\[2mm]
-&\frac{\i}{2}\left(a_{2}^{*}(\epsilon^{\12}\delta_{x})
-a_{2}(\epsilon^{\12}\delta_{x})\right)\left(a_{1}^{*}(\epsilon^{-\12}\delta_{x})
+a_{1}(\epsilon^{-\12}\delta_{x})\right)\\[2mm]
=&\frac{\i}{2}\left(a_{1}^{*}(\epsilon^{\12}\delta_{x})a_{2}(\epsilon^{-\12}\delta_{x})+
a_{1}^{*}(\epsilon^{-\12}\delta_{x})a_{2}(\epsilon^{\12}\delta_{x})\right.\\[2mm]
&-\left. a_{2}^{*}(\epsilon^{-\12}\delta_{x})a_{1}(\epsilon^{\12}\delta_{x}) -
a_{2}^{*}(\epsilon^{\12}\delta_{x})a_{1}(\epsilon^{-\12}\delta_{x})\right)\\[2mm]
+&\frac{\i}{2}\left(a_{1}^{*}(\epsilon^{\12}\delta_{x})a_{2}^{*}(\epsilon^{-\12}\delta_{x})
-a_{1}^{*}(\epsilon^{-\12}\delta_{x})a_{2}^{*}(\epsilon^{\12}\delta_{x})\right)\\[2mm]
+&\frac{\i}{2}\left( a_{2}(\epsilon^{\12}\delta_{x})a_{1}(\epsilon^{-\12}\delta_{x})
-a_{1}(\epsilon^{\12}\delta_{x})a_{2}(\epsilon^{-\12}\delta_{x})\right)\\[2mm]
=:&R^{a^{*}, a}(x)+ R^{a^{*},  a^{*}}(x)+ R^{a, a}(x).
\end{array}
\]
It is convenient to pass to the momentum representation using
the unitary  Fourier transform ${\mathcal F}$.
It follows that
\beq\label{e2.0}
\begin{array}{l}
a^{*}_{i}(\epsilon^{s}\delta_{x})=
(2\pi)^{-\12}\int_{\rr}\epsilon^{s}(k)\e^{-\i k x}a_{i}^{*}(k)\d
k,\\[2mm]
a_{i}(\epsilon^{s}\delta_{x})=
(2\pi)^{-\12}\int_{\rr}\epsilon^{s}(k)\e^{\i k x}a_{i}(k)\d
k.
\end{array}
\eeq
Let us first consider the term
\[
Q^{a^{*} \ a}(V)=\int_{\rr} V(x)R^{a^{*} \ a}(x)\d x.
\]
Using the above transformation and (\ref{e1.00}) we see that 
\beq
Q^{a^{*} \ a}(V)= \d\Gamma(\left[
\begin{array}{cc}
0&b\\b^{*}&0
\end{array}
\right]).
\label{e2.1}
\eeq
for
\beq
b= \frac{\i}{2}( \epsilon^{\12}V\epsilon^{-\12}+
\epsilon^{-\12}V\epsilon^{\12}).
\label{e2.2}
\eeq
Since $V, V'\in L^{\infty}(\rr)$, $V$ is a bounded operator on $H^{1}(\rr)$ hence
by interpolation  and duality  also on $H^{\12}(\rr)$ and
$H^{-\12}(\rr)$. This implies that $b$ is bounded. Clearly this
implies that (\ref{e2.0c}) holds for $Q^{a^{*}\ a}(V)$.
Let us now consider the term
\[
Q^{a^{*}\ a^{*}}(V)= \int_{\rr} V(x)R^{a^{*}\ a^{*}}(x)\d x.
\]
Using (\ref{e2.0}), we obtain that:
\[
\begin{array}{rl}
&Q^{a^{*}\ a^{*}}(V)\\[2mm]
=&\int_{\rr^{2}} R(k_{1}, k_{2})a_{1}^{*}(k_{1})a_{2}^{*}(k_{2})\d
k_{1}\d k_{2},
\end{array}
\]
where:
\beq\label{e2.0d}
R(k_{1}, k_{2})= \frac{\i}{4\pi}\widehat{V}(k_{1}+
k_{2})\left(\epsilon(k_{1})^{\12}\epsilon(k_{2})^{-\12}-
\epsilon(k_{1})^{-\12}\epsilon(k_{2})^{\12}\right).
\eeq
We note that:
\[
\begin{array}{rl}
&|\epsilon(k_{1})^{\12}\epsilon(k_{2})^{-\12}-
\epsilon(k_{1})^{-\12}\epsilon(k_{2})^{\12}|\\[2mm]
=&\epsilon(k_{1})^{-\12}\epsilon(k_{2})^{-\12}\left|\epsilon(k_{1})-
\epsilon(k_{2})\right|\\[2mm]
=&\epsilon(k_{1})^{-\12}\epsilon(k_{2})^{-\12}\left|\frac{k_{1}^{2}-k_{2}^{2}}{\epsilon(k_{1})+
\epsilon(k_{2})}\right|\\[2mm]
= &|k_{1}+ k_{2}|\left|\frac{k_{1}-k_{2}}{\epsilon(k_{1})+
\epsilon(k_{2})}\right|\epsilon(k_{1})^{-\12}\epsilon(k_{2})^{-\12}\\[2mm]
\leq &|k_{1}+
k_{2}|\epsilon(k_{1})^{-\12}\epsilon(k_{2})^{-\12}.
\end{array}
\]
Hence
\beq
|R(k_{1}, k_{2})|\leq C|\widehat{V'}(k_{1}+
k_{2})|\epsilon(k_{1})^{-\12}\epsilon(k_{2})^{-\12}.
\label{e2.3b}
\eeq
Arguing for example as in \cite{DG}, we obtain that
\beq
\|R\|_{L^{2}(\rr^{2})}\leq C \|V'\|_{L^{2}(\rr)}.
\label{e2.3}
\eeq
Using now the $N_{\tau}$ estimates (see \cite{GJ}), we obtain
(\ref{e2.0c}) for $Q^{a^{*} \ a^{*}}(V)$. The same estimate holds also for $Q^{a\
a}(V)$. \qed

\subsection{Coupling constant}
Let us fix a potential $V\in L^{\infty}(\rr)$ with $V'\in
L^{\infty}(\rr)\cap L^{2}(\rr)$. We set:
\beq\label{def-de-lambda}
(\lambda_{\rm quant})^{-1}:=\12\|\epsilon^{-1}V + V\epsilon
^{-1}\|_{B(L^{2}(\rr))}+
\frac{1}{m}\|\epsilon^{-\12}[V, \epsilon]\epsilon^{-\12}\|_{\rm HS}.
\eeq
\begin{lemma}\label{coupling-lemma}
Assume that $|\lambda|<\lambda_{\rm quant}$. Then:
\ben
\item there exists $0\leq \delta<1$ and $C\geq 0$ such that
\[
\pm \lambda Q(V)\leq \delta \d\Gamma(\omega)+ C.
\]
\item there exists $c>0$ such that
\[
\omega_{\lambda V}:=\mat{\epsilon}{\lambda b}{\lambda b^{*}}{\epsilon}\geq
c\one.
\]
\een
\end{lemma}
\proof Set $c_{0}=\|\epsilon^{-\12} b\epsilon^{-\12}\|=\12\|V \epsilon^{-1}+
\epsilon^{- 1}V\|$. Clearly 
\[
\pm \mat{0}{b}{b^{*}}{0}\leq c_{0}\mat{\epsilon}{0}{0}{\epsilon},
\]
hence
\[
\pm Q^{a^{*}\ a}(V)\leq c_{0}\d\Gamma(\omega).
\]
From the $N_{\tau}$ estimates (see eg \cite{GJ}) we get that
\[
\pm(Q^{a^{*}\ a^{*}}(V)+ Q^{a \ a}(V))\leq c_{1}(N+1)\leq
c_{1}m^{-1}(\d\Gamma(\omega)+ m),
\]
for
\[
c_{1}= 2\|R(\cdot, \cdot)\|_{L^{2}(\rr^{2})},
\]
for $R(k_{1}, k_{2})$ defined in (\ref{e2.0d}). Changing $k_{2}$ to
$-k_{2}$ and using (\ref{e1.00}), we see that
\[
c_{1}=\|\epsilon^{\12}V\epsilon^{-\12}-
\epsilon^{-\12}V\epsilon^{\12}\|_{\rm HS}= \|\epsilon^{-\12}[V,
\epsilon]\epsilon^{-\12}\|_{\rm HS}.
\]
These estimates clearly imply the lemma. 
\qed

\section{Charged $P(\varphi)_{2}$ Hamiltonians}\init\label{sec3}
In this section we construct the charged $P(\varphi)_{2}$ Hamiltonians
formally defined in (\ref{e1.003}). We also prove some resolvent
estimates known as {\em higher order estimates}.

\subsection{Charged $P(\varphi)_{2}$ Hamiltonians}
Let
\[
P(\lambda_{1},
\lambda_{2})=\sum_{|\alpha|=0}^{\rm deg P}a_{\alpha}
\lambda_{1}^{\alpha_{1}}\lambda_{2}^{\alpha_{2}}
\]
be a real  bounded below  polynomial on $\rr^{2}$. Clearly $P$ is  bounded below iff
${\rm deg }P=2m$ is even and 
\[
\inf_{\theta\in [0, 2\pi[}\sum_{|\alpha|= 2n}a_{\alpha}\cos
\theta^{\alpha_{1}}\sin \theta^{\alpha_{2}}>0.
\] 
Let also $g\in L^{2}(\rr)$ be a real function. We consider the
interaction term
\[
H_{I}= \int_{\rr}g(x):\! P(\varphi_{1}(x), \varphi_{2}(x))\!:\d x,
\]
where $\varphi_{i}(x)$ are defined in Subsect. \ref{sec2.1} and $:
\  \  :$ denotes the Wick ordering.

 By the usual arguments (see eg \cite{GJ},
\cite{DG}) one can show that $H_{I}$ is a {\em Wick polynomial}, ie a
finite sum of terms of the form:
\[
{\rm Wick}(w_{p, q})=\int_{\rr^{p+q}} w_{p, q}( k_{1}, \dots, k_{p},
k'_{1},\dots,
k'_{q})\prod_{1}^{p}a_{s_{i}}^{*}(k_{i})\prod_{1}^{q}
a_{r_{j}}(k'_{j})\d K \d K',
\]
where $s_{i}, r_{j}\in \{1, 2\}$ and the kernels $w_{p, q}$ are in
$L^{2}(\rr^{p+q})$.

Using the $N_{\tau}$ estimates  (see eg \cite{GJ, DG}) one can prove that
$H_{I}$ is a symmetric operator on $\Dom N^{m}$.

\begin{proposition}\label{3.1}
Assume that $g\in L^{2}(\rr)\cap L^{1}(\rr)$ and $g\geq 0$. Then
\ben
\item $H_{0}+ H_{I}$ is essentially selfadjoint on $\Dom H_{0}\cap
\Dom H_{I}$.

\item the operator $H_{1}= \overline{H_{0}+ H_{I}}$ is bounded below.
\item for any  $0<\epsilon$ there exists $C_{\epsilon}$ such that
\[
 H_{0}\leq (1+ \epsilon)H_{1}+ C_{\epsilon}.
\]
\een
\end{proposition}
\proof The proof is an immediate modification of arguments in the
standard $P(\varphi)_2$ model. One introduces the $Q-$space
representation associated to the canonical conjugation $F\mapsto
\overline{F}$ on $L^{2}(\rr; \cc^{2})$, which allows to identify
$\Gamma_{\s}(\ch)$ with $L^{2}(Q, \d\mu)$ for a probability measure
$\mu$. The operator $H_{I}$ can be
seen as a multiplication operator on $L^{2}(Q, \d\mu)$ such that
$H_{I}\in L^{p}(Q)$ for some $p>2$ and $\e^{- tV}\in L^{1}(Q)$ for
some $t>0$. To obtain the  second estimate one uses the fact that $g\geq 0$
and $P$ is bounded below. Using then that $\e^{- t H_{0}}$ is
hypercontractive, one obtains (1) and (2). The same argument show that
for any $\epsilon>0$ $\epsilon H_{0}+ H_{I}$ is bounded below, which
implies (3).  \qed

\medskip

The following {\em higher order estimates} are easily seen to hold  for $H_{1}$, 
with the same proof as in usual $P(\varphi)_{2}$ Hamiltonians.
\begin{proposition}\label{3.2}
Assume that $g\in L^{2}(\rr)\cap L^{1}(\rr)$ and $g\geq 0$. Then there exists 
$b>0$ such that for all $\alpha\in \nn$, the following 
{\em higher order estimates} hold:
\beq
\begin{array}{l}
\|N^{\alpha}(H_{1}+b)^{-\alpha}\|<\infty,\\[3mm]
\|H_{0}N^{\alpha}(H_{1}+b)^{-m-\alpha}\|<\infty,\\[3mm]
\|N^{\alpha}(H_{1}+b)^{-1}(N+1)^{1-\alpha}\|<\infty.
\end{array} 
\label{e3.1}
\eeq
\end{proposition}

\begin{theoreme}\label{3.3}
Assume that $g\in L^{2}(\rr)\cap L^{1}(\rr)$, $g\geq 0$, $V\in
L^{\infty}(\rr)$,  $V'\in L^{\infty}(\rr)\cap L^{2}(\rr)$.  

\ben
\item for any $\lambda$ with $|\lambda|<\lambda_{\rm quant}$, the
quadratic form $H_{0}+ \lambda Q(V)+ H_{I}$ with domain $\Dom
H_{0}\cap \Dom N^{m}$ is closeable and bounded below,
\item the domain of the closure of the above quadratic form equals
$\Dom |H_{1}|^{\12}$,
\item The associated bounded below selfadjoint operator will be
denoted by $H$ and called a {\em charged }$P(\varphi)_{2}$
{\em Hamiltonian}.
\een
\end{theoreme}
\proof From Lemma \ref{coupling-lemma}  we know that  if
$|\lambda|<\lambda_{\rm quant}$  then $\pm \lambda Q(V)\leq \delta
H_{0} +C$,  for some $0<\delta<1$.
By (3) of Prop. \ref{3.1}, this implies that as quadratic form $Q(V)$
is   $H_{1}-$ bounded with relative bound strictly less than $1$. The
theorem follows then from the KLMN theorem. \qed

\subsection{Higher order estimates and essential selfadjointness}
In this subsection we check that the higher order estimates of Prop.
\ref{3.2} extend to the full Hamiltonian $H$. As a consequence we will
find a suitable core for $H$.
\begin{proposition}\label{3.4}
Assume that $g\in L^{2}(\rr)\cap L^{1}(\rr)$, $g\geq 0$, $V\in
L^{\infty}(\rr)$,  $V'\in L^{\infty}(\rr)\cap L^{2}(\rr)$.  Let
$|\lambda|< \lambda_{\rm quant}$  and  $H$  the charged
$P(\varphi)_{2}$ Hamiltonian constructed  in Thm. \ref{3.3}. 
 Then there exists 
$b>0$ such that for all $\alpha\in \nn$, the following 
{\em higher order estimates} hold:
\beq
\begin{array}{l}
\|N^{\alpha}(H+b)^{-\alpha}\|<\infty,\\[3mm]
\|H_{0}N^{\alpha}(H+b)^{-m-\alpha}\|<\infty,\\[3mm]
\|N^{\alpha}(H+b)^{-1}(N+1)^{1-\alpha}\|<\infty.
\end{array} 
\label{e3.2}
\eeq
\end{proposition}
\begin{corollary}\label{th.10}
The Hamiltonian $H$ is essentially selfadjoint on $\Dom H_{0}\cap \Dom
N^{m}$. Consequently:
\[
H= (H_{0}+ \lambda Q(V)+ H_{I})^{\rm cl}= (\d\Gamma(\omega_{\lambda V})+ \lambda Q^{a^{*}\
a^{*}}(V) + \lambda Q^{a \ a}(V)+  H_{I})^{\rm cl},
\]
where we recall that:
\beq
\omega_{\lambda V}:= \left[\begin{array}{cc}
\epsilon&\lambda b\\\lambda b^{*}&\epsilon
\end{array}\right].
\label{e4.1}
\eeq
\end{corollary}
\proof It follows from Prop. \ref{3.4} that for $p$ large enough $\Dom
H^{p}\subset \Dom (H_{0})\cap \Dom N^{m}$. This implies the
corollary since $\Dom H^{p}$ is a core for $H$. \qed

\medskip

In the rest of this subsection we will explain the proof of Prop.
\ref{3.4}, which is 
 a rather  easy adaptation of the standard proof by Rosen
\cite{Ro}. We  will only
give the main steps, referring the reader for example to \cite [Sect. 7]{DG} for
details.

\medskip

{\it Lattices}

\medskip

The proof in \cite{Ro} relies on the introduction of a family $H_{n}$
of (volume and ultra-violet) cutoff Hamiltonians. These Hamiltonians are
obtained by considering an increasing sequence $\ch_{n}\subset \ch$
of finite dimensional subspaces of $\ch$ such that $\bigcup_{n\in
\nn}\ch_{n}$ is dense in $\ch$. Moreover one assumes that the
isometric projections $\pi_{n}: \ch\to \ch_{n}$ commute with the conjugation
$F\mapsto \overline{F}$ on $\ch= L^{2}(\rr; \cc^{2})$.

The subspaces $\ch_{n}$ are defined as follows: for $v\gg 1$, consider
the  lattice $v^{-1}\zz$ and let
\[
\rr\ni k\mapsto [k]_{v}\in v^{-1}\zz
\]
be the integer part mod $v^{-1}\zz$. For $\gamma\in v^{-1}\zz$, let
$e_{\gamma}(k)= v^{\12}\one_{]-(2v)^{-1}, (2v)^{-1}]}(k -\gamma)$. 
Set also for $\kappa\gg 1$
$\Gamma_{\kappa, v}= \{\gamma\in v^{-1}\zz \ : \
|\gamma|\leq\kappa\}$, and let
\[
\ch_{\kappa, v}:={\rm Span}\{e_{\gamma}\oplus 0, 0\oplus e_{\gamma}
\ : \gamma\in \Gamma_{\kappa, v}\}.
\]

We choose then a sequence $(\kappa_{n}, v_{n})$ tending to $(\infty,
\infty)$ in such a way that $\Gamma_{\kappa_{n}, v_{n}}\subset
\Gamma_{\kappa_{n+1}, v_{n+1}}$ and set $\ch_{n}:= \ch_{\kappa_{n},
v_{n}}$.

\medskip

{\em Cutoff Hamiltonians}

\medskip

Let us explain how to define the associated cutoff Hamiltonians. Since
$\ch= \ch_{n}\oplus \ch_{n}^{\perp}$, there exists by the exponential
law a unitary map $U_{n}: \Gamma_{\s}(\ch_{n})\otimes
\Gamma_{\s}(\ch_{n}^{\perp})\to
\Gamma_{\s}(\ch)$. If $W$ is a bounded operator on $\Gamma_{\s}(\ch)$,
one can define its projection to $\Gamma_{\s}(\ch_{n})$:
\beq\label{th.2}
\Pi_{n}W:=
U_{n}\left(\Gamma(\pi_{n})W\Gamma(\pi_{n})^{*}\otimes\one\right)U_{n}^{-1}.
\eeq
This definition extends to Wick polynomials, for example if 
$W= \prod_{1}^{p}a^{*}(F_{i})\prod_{1}^{q}a(G_{i})$, then:
\[
\Pi_{n}W=\prod_{1}^{p}a^{*}(\pi_{n}^{*}\pi_{n}F_{i})\prod_{1}^{q}a(\pi_{n}^{*}\pi_{n}G_{i}).
\]
We set now:
\[
H_{0,
n}:=\d\Gamma(\epsilon_{n}\oplus \epsilon_{n}),\ H_{I, n}:=
\Pi_{n}H_{I}, \ Q_{n}(V):= \Pi_{n}Q(V),
\]
 where $\Pi_{n}W$ is defined in (\ref{th.2}) and 
\[
\epsilon_{n}(k)= \epsilon([k]_{v_{n}}),
\]
in the momentum space representation. Note that $\epsilon_{n}\oplus \epsilon_{n}$ commutes with
$\pi_{n}^{*}\pi_{n}$.  The construction of the cutoff Hamiltonians
$H_{n}$ is done in the next proposition.
\begin{proposition}\label{th.11}
\ben
\item Let $|\lambda|< \lambda_{\rm quant}$.  Then there exists
$0<\delta<1$  and $C>0$ such that uniformly for $n$ large enough:
\[
\pm \lambda Q_{n}(V)\leq\delta H_{0, n}+C . 
\]
\item the Hamiltonian $H_{1, n}=H_{0, n}+ H_{I, n}$ is essentially
selfadjoint on $\Dom\d\Gamma(\omega)\cap \Dom
N^{m}$ and there exists $b>0$ such that
\[
0\leq H_{1, n}+b, \ \forall \ n \in \nn.
\]
\item there exists $0<\delta<1$  and $b>0$ such that
\[
\pm \lambda Q_{n}(V)\leq \delta (H_{1, n}+ b), \ \forall \ n\in \nn.
\]
\item Let $H_{n}$ the bounded below selfadjoint operator associated to
the quadratic form $H_{1, n}+ \lambda Q_{n}(V)$ with domain
$\Dom|H_{1, n}|^{\12}$. Then
\[
\slim_{n\to \infty} (H_{n}+ b)^{-1}= (H+ b)^{-1},
\]
where $H$ is the charged $P(\varphi)_{2}$ Hamiltonian defined in Thm.
\ref{3.3}.
\een
\end{proposition}
To prove (1) we note that (modulo the trivial factors $U_{n}$):
\beq\label{th.01}
\Pi_{n}W= \Gamma(\pi_{n}^{*}\pi_{n})W\Gamma( \pi_{n}^{*}\pi_{n}).
\eeq
Since $|\lambda|< \lambda_{\rm quant}$ there exists $0<\delta<1$ and
$C>0$ such
that  $\lambda Q(V)\leq\delta \d\Gamma(\epsilon\oplus \epsilon)+C$. Using
(\ref{th.01}) we get that
\[
\lambda Q_{n}(V)\leq \delta\Gamma(\pi_{n}^{*}\pi_{n})\d\Gamma(\epsilon\oplus
\epsilon)\Gamma(\pi_{n}^{*}\pi_{n})+C\leq \delta\d\Gamma(\epsilon\oplus
\epsilon)+C, 
\]
since $\pi_{n}^{*}\pi_{n}$ commutes with $\epsilon\oplus \epsilon$. Clearly for any $\alpha>0$ one has
\[
(1+\alpha)^{-1}\epsilon_{n}\leq \epsilon\leq (1+
\alpha)\epsilon_{n}, \hbox{ if }n\hbox{ is large enough}.
\]
This implies (1). Statement (2) is  standard (see eg
\cite[Sect. 7]{DG}). It follows also that
for any $\epsilon>0$ there exists $C_{\epsilon}$ such that
uniformly in $n$:
\[
H_{0, n}\leq (1+ \epsilon)H_{1, n}+ C_{\epsilon},
\]
which implies (3). It remains to prove (4).
Since $Q_{n}(V)$  are uniformly $H_{1,n}-$ form
bounded with relative bound strictly less than $1$, there exists
$b\gg 1$ such that $(H_{1, n}+ b)^{-\12}\lambda Q_{n}(V)(H_{1, n}+ b)^{-\12}$
 has norm 
less than some $\delta<1$ uniformly in $n$, and:
\[
(H_{n}+ b)^{-1}= (H_{1, n}+ b)^{-\12} (\one +  (H_{1, n}+
b)^{-\12}\lambda Q_{n}(V)(H_{1, n}+ b)^{-\12})^{-1}(H_{1, n}+ b)^{-\12},
\]
It follows that 
\[
(H_{n}+ b)^{-1}= \sum_{k=0}^{+\infty}(H_{1, n}+
b)^{-1}(-\lambda Q_{n}(V)(H_{1, n}+ b)^{-1})^{k}
\]
as a norm convergent series. The same formula holds for $(H+b)^{-1}$.
Therefore it suffices to prove that for all $k\in \nn$:
\beq
\slim_{n\to \infty}(H_{1, n}+
b)^{-1}(Q_{n}(V)(H_{1, n}+ b)^{-1})^{k}=(H_{1}+ b)^{-1}(Q(V)(H_{1}+
b)^{-1})^{k}.
\label{th.4}
\eeq
 The arguments in \cite[Prop. 4.8]{SHK}
easily extend to yield  that 
\beq\label{th.5}
(H_{1, n}+ b)^{-1}\to (H_{1}+b)^{-1}\hbox{ in norm}.
\eeq
(Note that $H_{1}$ is a essentially a standard $P(\varphi)_{2}$
Hamiltonian).
Moreover
\beq
\sup_{n\in \nn}\|N(H_{1, n}+ b)^{-1}\|<\infty.
\label{th.6}
\eeq 
This implies using  Lemma \ref{2.1} that $Q_{n}(V)(H_{1,
n}+ b)^{-1}$ is uniformly bounded. Hence (\ref{th.4}) will follow from
\beq
\slim_{n\to \infty}(H_{1}+
b)^{-1}(Q_{n}(V)(H_{1}+ b)^{-1})^{k}=(H_{1}+ b)^{-1}(Q(V)(H_{1}+
b)^{-1})^{k}.
\label{th.6b}
\eeq
Now $Q_{n}(V)(H_{1}+b)^{-1}$ is uniformly bounded and converges
strongly to $Q(V)(H_{1}+ b)^{-1}$, which implies (\ref{th.6b}). This
completes the proof of the proposition.\qed

\medskip

{\bf Proof of Prop. \ref{3.4}.}
The key  point of the proof of the higher order estimates is to
consider the multicommutators:
\[
R_{i, n}(k_{1}, \dots, k_{p}):= {\rm ad}_{a_{i}(k_{1})}\cdots {\rm
ad}_{a_{i}(k_{p})}(H_{I,n}+ 
Q_{n}(V)), \ k_{1},\dots, k_{p}\in \rr,
\]
for $i=1, 2$ where ${\rm ad}_{A}B = [A, B]$.  The key step is then to
prove that there exists $b>0$ such that for all $\lambda_{1},
\lambda_{2}\geq b$ one has:
\beq
\|(H_{n}+ \lambda_{1})^{-\12}R_{i,n}(k_{1}, \dots, k_{p})(H_{n}+
\lambda_{2})^{-\12}\|\leq C_{p}\prod_{1}^{p} F_{n}(k_{i}),
\label{e3.4}
\eeq
where 
\beq\label{e3.4b}
\sup_{n\in \nn}\int_{\rr}|F_{n}(k)|^{2}\epsilon(k)^{-\delta}\d k<\infty, \ \forall \ 
\delta>0.
\eeq

Note  that it is only
necessary to bound multicommutators with $a_{i}(k)$ for a {\em fixed}
$i=1, 2$. Indeed this follows from the fact that it suffices to prove
the higher order estimates with $N$,  $H_{0}$ replaced by $N_{i}$,
$H_{0,i}$ for:
\[
 N_{i}=\int_{\rr}a_{i}^{*}(k)a_{i}(k)\d k, \ H_{0,
i}=\int_{\rr}\epsilon(k)a_{i}^{*}(k)a_{i}(k)\d k.
\]
We first note that since $H_{0,n}\leq C(H_{n}+ b)$ uniformly in $n$, it
suffices to prove (\ref{e3.4}) with $(H_{0,n}+ \lambda)^{-\12}$ instead
of $(H_{n}+ \lambda)^{-\12}$.

Clearly  the multicommutator 
$R_{i}(\cdots)$ is the sum of the two multicommutators with $H_{I, n}$
and $Q_{n}(V)$. The multicommutators with $H_{I,n}$ are estimated as in
\cite{Ro}, \cite {DG}, yielding:
\beq
\|(H_{0}+ \lambda_{1})^{-\12}{\rm ad}_{a_{i}(k_{1})}\cdots {\rm
ad}_{a_{i}(k_{p})}H_{I,n}(H_{0}+
\lambda_{2})^{-\12}\|\leq C_{p}\prod_{1}^{p} \epsilon(k_{i})^{-\12},
\label{e3.5}
\eeq
so that (\ref{e3.4b}) is satisfied.  

Let us now estimate the multicommutators with $Q_{n}^{a^{*}\ a}(V)$.
 Abusing notation, we will still
denote by $\pi_{n}$ the projection from $L^{2}(\rr)$ onto ${\rm
Span}\{e_{\gamma}\ : \ \gamma\in \Gamma_{\kappa_{n}, v_{n}}\}$. Let $b_{n}=\pi_{n}^{*}\pi_{n}b \pi_{n}^{*}\pi_{n}$, where $b$ is
the operator defined in (\ref{e2.2}) and denote also by $b_{n}(k_{1},
k_{2})$ its kernel in the momentum representation.  Then
\[
\ad_{a_{1}(k)}Q^{a^{*}\ a}_{n}(V)= a_{2}(b_{n}(k, \cdot)), 
\]
and the similar formula with the indices $1$ and $2$ exchanged. 
 Using
the well known estimate
\beq\label{e3.7}
\|a(f)(\d\Gamma(b)+1)^{-\12}\|\leq \|b^{-\12}f\|,
\eeq
for $b\geq 0$, we get
\[
\|\ad_{a_{1}(k)}Q^{a^{*}\ a}_{n}(V)(H_{0, n}+ b)^{-\12}\|\leq
\|\epsilon_{n}(\cdot)^{-\12} b_{n}(k, \cdot)\|_{L^{2}(\rr)}=: F_{n}(k), 
\]
hence to prove (\ref{e3.4b}) it suffices to show that 
$\epsilon(k_{1})^{-\delta/2}\epsilon(k_{2})^{-\12}b_{n}(k_{1},
k_{2})\in L^{2}(\rr^{2})$ uniformly in $n$. This is equivalent to the fact that
$\epsilon^{-\delta/2}b_{n}\epsilon^{-\12}$  is Hilbert-Schmidt
uniformly in $n$. 
Clearly this is true if $\epsilon^{-\delta/2}b\epsilon^{-\12}$ is
Hilbert-Schmidt. Working in the momentum representation we  
need to consider the
integrals:
\[
I_{1}=\int \epsilon(k)^{-1- \delta}|\widehat{V}|^{2}(k'- k)\d k \d k',
\]
\[
I_{2}= \int\epsilon(k)^{1- \delta} \epsilon(k')^{-2}|\widehat{V}|^{2}(k'-
k)\d k \d k'.
\]
$I_{1}$ is clearly convergent since $V\in L^{2}(\rr)$. To estimate
$I_{2}$, we use the Peetre inequality:
\beq
1+ |x|\leq 2(1+ |x-y|)(1+  |y|), \ x, y\in \rr,
\label{e3.6}
\eeq
and obtain that $I_{2}$ is convergent.  In fact
$\epsilon(k)^{(1-\delta)/2}\widehat{V}\in L^{2}(\rr)$ since $V'\in L^{2}(\rr)$.

Let us now consider the multicommutators with $Q^{a^{*}\
a^{*}}_{n}(V)$. Recall that the kernel $R(k_{1}, k_{2})$ of $Q^{a^{*}\
a^{*}}(V)$ was defined in (\ref{e2.0d}) and set $R_{n}=
\Gamma(\pi_{n}^{*}\pi_{n})R$. Then:
\[
\ad_{a_{1}(k)}Q^{a^{*}\ a^{*}}_{n}(V)= a_{2}^{*}(R_{n}(k, \cdot)).
\]
Using again (\ref{e3.7}), we get that
\[
\|(H_{0, n}+ b)^{-\12}\ad_{a_{1}(k)}Q_{n}^{a^{*}\ a^{*}}(V)\|\leq
\|\epsilon_{n}(\cdot)^{-\12}R_{n}(k, \cdot)\|_{L^{2}(\rr)}=: F_{n}(k).
\]
Now(\ref{e3.4b}) follows from the fact that $R(k_{1}, k_{2})\in
L^{2}(\rr^{2})$,shown in (\ref{e2.3}). 

The proof of the higher order estimates can now be completed as in
\cite{Ro}, \cite{DG}.  In particular the strong resolvent convergence
in Prop. \ref{th.11} (4) is needed to apply the principle of cutoff
independence in \cite{Ro}. \qed

\section{Spectral and scattering theory for charged $P(\varphi)_{2}$
Hamiltonians}\init\label{sec4}
In this section we study the spectral and scattering theory of charged $P(\varphi)_{2}$
Hamiltonians. We will use the  results of \cite{GP}. In
\cite{GP}, we introduced an abstract class of QFT Hamiltonians formally
given by 
\[
H= \d\Gamma(\omega)+ {\rm Wick}(w),
\]
on a bosonic Fock space $\Gamma_{\s}(\ch)$, where $\omega\geq 0$ is a
selfadjoint operator on the one-particle space $\ch$ and ${\rm
Wick}(w)$ is a  Wick polynomial associated to a kernel $w$.

Our main task in this section will be to explain how to fit charged $P(\varphi)_{2}$
Hamiltonians into the abstract framework of \cite{GP} and to check the
various abstract hypotheses there. The results on spectral and
scattering theory are then obtained as simple applications of the
generals results of \cite{GP}.
\subsection{Charged $P(\varphi)_{2}$ Hamiltonians 
as abstract QFT Hamiltonians}\label{sec4.1}
The class of abstract QFT Hamiltonians in \cite{GP} is described in
terms of three types of hypotheses, which will be briefly explained
below.

\medskip

{\it Hypotheses on the Hamiltonian}

\medskip

One  first requires (see \cite[Subsect. 3.1]{GP}) that  the Hamiltonian $H$ is the closure of
$\d\Gamma(\omega)+ {\rm Wick}(w)$ where ${\rm Wick}(w)$ is a Wick
polynomial with $L^{2}$ kernels and is bounded below.
This follows from Corollary \ref{th.10}. In our case we take for
$\omega$  the operator $\omega_{\lambda V}$ 
defined in (\ref{e4.1}). 

One also requires that $\omega\geq m_{1}>0$, which follows from Lemma
\ref{coupling-lemma}.

 Moreover one
asks that any power of the number operator should be controlled by a
sufficiently high power of the resolvent of $H$ (see
\cite[Subsect. 3.1]{GP}). This follows from the higher order
estimates, which were proved in Prop. \ref{3.4}.

\medskip

{\it Hypotheses on the one-particle Hamiltonian}

\medskip

On requires that the one-particle energy $\omega$ has a sufficiently
nice spectral and scattering theory. The precise statements can be
found in \cite[Subsect. 3.2]{GP}. They are formulated in terms of two
additional selfadjoint operators on the one-particle Hilbert space
$\ch$. 

The first one, denoted by $\langle X\rangle$ is  called a {\em
weight operator}, used to measure the propagation of one-particle
states to infinity. The second, denoted by $a$ is a {\em conjugate
operator}, used in the Mourre commutator method. Moreover one
introduces a dense subspace ${\mathcal S}$ of $\ch$, preserved by the
operators $\omega, a, \langle X\rangle$ on which (multi)-commutators
between these three operators can be unambiguously defined.

To verify them in our case it is
convenient to assume that the electrostatic potential is smooth. More
precisely we will assume that $V\in S^{- \mu}(\rr)$, for some $\mu>0$, where the classes
$S^{m}(\rr)$ are defined in Subsect. \ref{sec0.3}. 

The one-particle
Hamiltonian in our case is $\omega_{\lambda V}$ defined in  (\ref{e4.1}).
For the weight operator, we choose:
\[
\langle X\rangle:=\mat{\langle x\rangle}{0}{0}{\langle x\rangle}, 
\]
and for the conjugate operator
\[
a=\mat{c}{0}{0}{c}, \ c= \12(x\cdot\frac{D_{x}}{ \epsilon(D_{x})}+
\frac{D_{x}}{ \epsilon(D_{x})}\cdot x).
\]
For the subspace ${\mathcal S}$ we choose ${\mathcal S}(\rr)\oplus{\mathcal
S}(\rr)$ where 
${\mathcal S}(\rr)$ is  the Schwartz class.

Assuming that  $V\in S^{-\mu}(\rr)$ for some $\mu>0$, it is a tedious
but straightforward exercise in pseudodifferential calculus to check
that the hypotheses in \cite[Subsect. 3.2]{GP} are satisfied.

\medskip

{\it Hypotheses on the interaction}

\medskip

The final set of hypotheses concerns the kernel $w$ of the interaction
${\rm Wick}(w)$ (see \cite[Subsect. 3.3]{GP}). In our case they
correspond to the fact  that each
kernel $w_{p, q}$, considered as an element of
$\otimes^{p+q}L^{2}(\rr; \cc^{2})$ 
should be in  the domain of $\d\Gamma(\langle x\rangle^{s})$ for some
$s>1$. 

The interaction term ${\rm Wick(w)}$ is the sum of the
$P(\varphi)_{2}$ interaction $\int_{\rr} :\! P(\varphi_{1}(x),
\varphi_{2}(x))\!:\d x$ and  $\lambda
Q^{a^{*}\ a^{*}}(V)+ \lambda Q^{a\ a}(V)$.

For the first term, the hypotheses above are satisfied if $\langle
x\rangle^{s}g\in L^{2}(\rr)$ (see \cite[Subsect. 6.3]{DG}). 

\begin{lemma}\label{4.1}
Assume that $V\in S^{-\mu}(\rr)$ for some $\mu>\12$. Let $R(k_{1},
k_{2})$ be the kernel of $Q^{a^{*}\ a^{*}}(V)$ and $Q^{a\ a}(V)$
defined in (\ref{e2.0d}). Then $|D_{k_{i}}|^{s}R\in L^{2}(\rr^{2})$
for $i=1, 2$.
\end{lemma}
\proof 
Using (\ref{e1.00}), we see that $R(k_{1}, -k_{2})$ is the
distribution kernel of 
\[
\begin{array}{rl}
&\frac{\i}{2}{\mathcal F}( \epsilon(D_{x})^{\12}V\epsilon(D_{x})^{-\12}-
\epsilon(D_{x})^{-\12}V \epsilon(D_{x})^{\12}){\mathcal F}^{-1}\\[2mm]
=& 
\12 {\mathcal F}( \epsilon(D_{x})^{-\12}[\epsilon(D_{x}), \i
V]\epsilon(D_{x})^{-\12}{\mathcal F}^{-1}.
\end{array}
\]
We need to prove that $|D_{k_{1}}|^{s}R( k_{1}, -k_{2})\in
L^{2}(\rr^{2})$ or equivalently that  the operator $C=\langle
x\rangle^{s}\epsilon(D_{x})^{-\12}[\epsilon(D_{x}),
V]\epsilon(D_{x})^{-\12}$ is Hilbert Schmidt on $L^{2}(\rr)$.

From the pseudodifferential calculus, we obtain that  $C= \Op^{\rm
w}(c)$, where $c(x, k)$  is a symbol satisfying:
\[
|\p_{x}^{\alpha}\p_{k}^{\beta}c(x, k)|\leq C_{\alpha, \beta}\langle
x\rangle^{-\mu -1+ s}\langle k\rangle^{-1- \beta}, \ \alpha, \beta\in\nn,
\]
and $\Op^{\rm w}(a)$ denotes the Weyl quantization of $a$.
Using (\ref{e1.0a}) we see  the conclusion of the lemma holds if  $\mu
>\12$. \qed

\subsection{Spectrum of charged $P(\varphi)_{2}$ Hamiltonians}
In the rest of this section we assume:
\[
\langle x^{s}\rangle g\in L^{2}(\rr), , \ g\in
L^{1}(\rr), \ g\geq 0,\ V\in S^{-s}(\rr), \hbox{ for  some }s>1.
\]
Moreover we assume as before that:
\[
|\lambda|<\lambda_{\rm quant}.
\]
\begin{theoreme}[HVZ Theorem]
One has
\[
\sigma_{\rm ess}(H)= [\inf \sigma(H)+ m, +\infty[.
\]
Consequently $H$ has a ground state.
\end{theoreme}
The theorem follows from \cite[Thm. 7.1]{GP} and the fact that
$\sigma_{\rm ess}(\omega_{\lambda V})= [m, +\infty[$.

\subsection{Asymptotic fields}
For $F\in \ch$ we set $F_{t}:=\e^{-\i t \omega}h$.  The results of
this subsection follow from \cite[Thm. 4.1]{GP}, taking into account
\cite[Remark 4.2]{GP}. The fact that $\omega_{\lambda V}$ restricted to its
continuous spectral subspace is unitarily equivalent to $\omega$
follow easily from standard two-body scattering theory, using that
$V\in S^{-s}(\rr)$ for $s>1$. 
\begin{theoreme}
\ben  
\item for all $F\in \ch$ the strong limits
\beq
W^{\pm}(F):= \slim_{t\rightarrow \pm\infty}\e^{\i tH}W (F_{t})\e^{-\i tH}
\label{eas.3bis}
\eeq
exist. They are called the {\em asymptotic Weyl operators}.
\item 
the map
\beq\ch\ni F\mapsto W^{\pm}(F)\label{kwer.1}\eeq
is strongly continuous.
\item the operators $W^{\pm}(F)$ satisfy the Weyl commutation relations:
\[
W^{\pm}(F)W^{\pm}(G)= \e^{-\i\12{\rm Im}(F|G)}W^{\pm}(F+G).
\]
\item the Hamiltonian preserves the asymptotic Weyl operators:
\beq
\e^{\i tH}W^{\pm}(F)\e^{-\i tH}= W^{\pm}(F_{-t}).
\label{eas.3er}
\eeq
\een
\label{4.1bis}
\end{theoreme}

\subsection{Wave operators and asymptotic completeness}
For $F\in \ch$, let 
$a^{\pm\sharp}(F)$ the asymptotic creation/annihilation operators
associated to the asymptotic Weyl operators (see eg \cite[Subsect.
8.1]{GP}). The following theorem describes the construction of {\em
wave operators} and their main property, the {\em asymptotic
completeness}.
\begin{theoreme}
Set:
\[
\begin{array}{rl}
\Omega^{\pm}:& \cH_{\rm pp}(H)\otimes \Gamma_{\s}(\ch)\to \Gamma_{\s}(\ch)\\[2mm]
&\Psi\otimes\prod_{1}^{n}a^{*}(F_{i})\Omega\mapsto \prod_{1}^{n}a^{\pm
*}(F_{i})\Psi.
\end{array}
\]
The operators $\Omega^{\pm}$ are called the {\em wave operators}.
Set also
\[
H^{\pm}= H_{|\cH_{\rm pp}(H)}\otimes\one + \one\otimes \d\Gamma(
\omega),\hbox{ acting on }\cH_{\rm pp}(H)\otimes \Gamma_{\s}(\ch).
\]
The operators $H^{\pm}$ are called the {\em asymptotic Hamiltonians}.
Then:
\ben
\item $\Omega^{\pm}$  are unitary operators;
\item $\Omega^{\pm}$ intertwine the asymptotic Weyl operators with the
Fock Weyl operators:
\[
\Omega^{\pm}\one\otimes W(F)=  W^{\pm}(F)\Omega^{\pm}, \ \forall \ F\in
\ch,
\]
\item $\Omega^{\pm}$ intertwine the asymptotic Hamiltonians  with the
Hamiltonian $H$:
\[
\Omega^{\pm}H^{\pm}=  H\Omega^{\pm}.
\]
\een
\end{theoreme}

\end{document}